\documentstyle[a4,epsfig]{report}   
\def\ZzZ{{\hbox{\tenrm Z\kern-.31em{Z}}}}   
\def\CcC{{\hbox{\tenrm C\kern-.45em{\vrule height.67em width0.08em depth-   
.04em   
\hskip.45em }}}}   
\def\mapright#1{\smash{\mathop{\longrightarrow}\limits^{#1}}}   
\def\mapbelow#1{\smash{\mathop{\longrightarrow}\limits_{#1}}}   
   
\newcommand{\ep}{\epsilon}   
\newcommand{\lab}{\label}

\newcommand{\bc}{\begin{center}}   
\newcommand{\ec}{\end{center}}   
\newcommand{\be}{\begin{equation}}   
\newcommand{\ee}{\end{equation}}   
\newcommand{\bea}{\begin{eqnarray}}   
\newcommand{\eea}{\end{eqnarray}}   
\newcommand{\bs}{\begin{subequations}}   
\newcommand{\es}{\end{subequations}}   
\newcommand{\beq}{\begin{eqalignno}}   
\newcommand{\eeq}{\end{eqalignno}}   
\def\bol#1{\mbox{\boldmath\tiny $#1$\normalsize\unboldmath}}   
\def\vec#1{\mbox{\boldmath $#1$\unboldmath}}   
\def\bol#1{{\bf #1}}   
\def\vec#1{{\bf #1}}

\newcommand{\half}{\frac{1}{2}}   
\newcommand{\qrt}{\frac{1}{4}}   
   
\def\Journal#1#2#3#4{{#1} {\bf #2}, {#3} {(#4)}}   
   
\def\PRD{{\em Phys. Rev.}  \bf D}   
   
\def\om{\omega}   
\def\Om{\Omega}   
   
\def\lab{\label}   
\begin{document}   
%
   
  
\vspace{2.0cm}  
\bc  
\Large{\bf Vacuum structure for expanding geometry}    
  
\vspace{1.2cm}  
  
\large{ E. Alfinito$^{1, 3}$, R. Manka$^{4}$ and G. Vitiello$^{1,2}$} \\  
\small  
{ \it ${}^{1}$Dipartimento di Fisica, Universit\`a di Salerno, 84100   
Salerno, Italy\\ 
 ${}^{2}$ INFN Gruppo Collegato di Salerno\\  
 ${}^{3}$ INFM Sezione di Salerno}\\ 
{\it ${}^{4}$Physics Department, University of Katowice, Katowice,   
Poland}  
\vspace{1.5cm}

\ec  
\normalsize  
{\bf Abstract} We consider gravitational wave modes in the   
FRW metrics in a  
 de Sitter    
phase and show that the state space splits into many unitarily    
inequivalent representations of the canonical commutation relations.   
Non-unitary time evolution    
is described as a trajectory in the space    
of the representations. The generator of time evolution    
is related to the entropy operator. The thermodynamic arrow of  
time is shown to point in the same direction of the cosmological  
arrow of time. The vacuum is a two-mode    
SU(1,1) squeezed state of    
thermo field dynamics. The link between    
expanding geometry, squeezing and thermal properties   
is exhibited.  
   
\vspace{2cm}

\normalsize

\newpage

\setcounter{chapter}{1}   
\setcounter{equation}{0}   
\section*{1. Introduction}   
   
Expanding and inflationary Universe scenarios have gained a central    
position in the interest of cosmologists and, more generally, of    
physicists interested in quantum aspects    
of General Relativity. In fact, a phase of    
primordial inflation explains (or seems to explain) some open problems in    
Standard Cosmology such as the flatness and horizon problems   
and it also sheds some light on the origin and on the r\"ole of   
quantum fluctuations \cite{1, GU, 2}.    
It is therefore of    
crucial relevance the possibility of describing in quantum field theory    
(QFT) the non-unitary time evolution implied by expanding metrics    
models.   
   
As well known, the same concept of {\it vacuum} state   
is meaningless in the presence of a    
curved space-time background. The particle production in a gravitational    
field \cite{Park}, the Hawking radiation \cite{5} and the generation of    
gravitational waves from vacuum in an expanding Universe \cite{Grish1,    
Grish2, Grish3} are indeed strictly related with the problem of properly    
defining the particle number operator in curved   
space-time (see also \cite{MSV} and \cite{WA1, WA2}).   
   
Even more serious is the problem of quantization   
in the expanding Universe, since to the problem of the    
vacuum definition in    
curved space-time adds up the problem of the non-unitary   
time evolution. In this paper we focus our   
study on the problem of the quantization of gravitational modes   
in expanding geometry in the canonical operator formalism. 
We do not consider the case of the   
free and self-interacting scalar matter field which has been   
considered elsewhere, see e.g. \cite{Guth, Eboli, Allen}, also in   
the presence   
of black hole solutions \cite{MSV}.  
  
We will see that the whole set of unitarily inequivalent   
representations of the canonical commutation relations   
will be required in our discussion. Our result shows that    
the system state space splits into many unitarily    
inequivalent representations, each one labeled by different time    
values.    
Time evolution is described as a trajectory in the space of the    
representations: the system evolves in time by running over unitarily    
inequivalent representations. The non-unitary character of time    
evolution implied by expanding metrics thus finds its description in the   
non-unitary equivalence among the representations at    
different $t$'s.   
We show that the generator responsible    
for non-unitary time evolution is related to the entropy operator,    
which is indeed consistent (see \cite{DFV, QD}) with the fact that    
inflation implies irreversibility in time evolution ({\it the arrow of    
time}). In this connection, we will see how our result   
relates to some of the features of the arrows of time which have been   
discussed by  
Penrose and Hawking, although from different perspectives and on   
the basis of semi-classical arguments \cite{Penrose, Hawking, book}  
(see   
also the related discussion in \cite{Elitzur}). We find that, also 
consistently with   
Penrose and Hawking discussions, 
our quantum formalism implies that the   
thermodynamic arrow of time (the direction of time in which   
entropy increases) is in accordance with the cosmological arrow   
of time (the direction of time in which the Universe expands).  
  
The system states are recognized to be    
SU(1,1) squeezed coherent states of the same kind of those    
of Thermo Field Dynamics   
(TFD) \cite{TFD, Um1, Um} (the TFD formalism anyway    
underlies many works   
since the paper by Israel \cite{Israel} on TFD of black    
holes; see also \cite{Jo} and for rigorous results in the algebraic    
approach see \cite{SW}).   
   
The emerging picture of    
the canonical quantization scheme we obtain is thus a    
unified view of many    
features of time evolution in expanding geometry, including coherence,    
two-mode squeezing, entropy and vacuum thermal properties    
\cite{Grish1, Grish2, Grish3, Prok, BraProk, Albr}.

The paper is organized as follows. In Sect. 2 we introduce general,    
well known features of the quasi-linear approximation for the   
Friedmann-Robertson-Walker (FRW)    
expanding metrics case.   
In Sect. 3 we start the discussion of the system    
quantization.   
In Sect. 4 we obtain the    
Hamiltonian spectrum by the method of the spectrum generating algebra and    
exhibit the theory vacuum structure, its time evolution and its two-mode    
squeezing character.    
In Sect. 5    
we discuss the evolution as trajectories    
over unitarily inequivalent representations. Entropy, free energy and    
the first principle of thermodynamics are discussed in Sect. 6.  
The infrared cut-off which is present in  
our formalism is considered in Sect. 7 where the emerging  
of a life-time for each $k$-mode is also  
shown. Concluding remarks are reported in Sect. 8.  
In this paper we do not consider specific    
model features, or renormalization    
problems, neither we study symmetry restoration mechanisms due to thermal    
effects.   
Details of the mathematical  
formalism are presented in the Appendices.

\setcounter{chapter}{2}   
\setcounter{equation}{0}   
\section*{2. Expanding Universe}   
   
In the four dimensional space-time $x^{\mu} = \{x_{0}= ct, x^{i}\},\,    
i=1,2,3, $ we consider the   
linear approximation in the weak-field limit 
$|h_{\mu\nu}| << 1$ \cite{MTW}. One decomposes the metrics    
$g_{\mu\nu}$ as   
\be g_{\mu\nu}\, =\,g^{0}_{\mu\nu}\,+\,h_{\mu\nu}.     
\lab{p1}\ee   
When one chooses the flat background metrics   
\be    
g_{\mu \nu }^0=\eta _{\mu \nu }=\left(     
\begin{tabular}{llll}    
1 & 0 & 0 & 0 \\     
0 & -1 & 0 & 0 \\     
0 & 0 & -1 & 0 \\     
0 & 0 & 0 & -1    
\end{tabular}    
\right)  ~~,   
\lab{p2}\ee   
as customary one defines   
\be   
\overline{h}_{\mu \nu }=h_{\mu \nu }-\frac 12\eta _{\mu \nu }h     
~~, \lab{p3}\ee   
with $h\,=\, h^{\mu}_{\mu}$. The (de Donder) gauge condition   
$ \partial _\mu \overline{h}^{\mu \nu }=0$ and     
the Einstein equations   
give   
\be   
\Box \overline{h}^{\mu \nu }=0.     
\lab{p9}\ee   
The field $\overline{h}^{\mu \nu }$ is then decomposed as usual   
\cite{MTW} into partial waves   
\be   
\overline{h}^{\mu \nu }=\sum_\lambda \frac 1{(2\pi )^{\frac 32}}\int    
d^3 {\vec k}e_{{\vec k}\lambda }^{\mu \nu }\{u_{{\bol k}\lambda }(t)   
e^{i{\bol k}\cdot     
{\bol x}}+u_{{\bol k}\lambda }^{\dag}(t)e^{-i{\bol k}\cdot{\bol x}   
}\}      
\lab{p10}\ee   
with    
$k\equiv(k_0=\omega =ck,{\vec k})$ . The wave function   
$u_{{\bol k}\lambda }(t)$     
satisfies the    
simple harmonic oscillator equation   
\be   
\frac{d^2}{dt^2}u_{{\bol k}\lambda}(t)+\omega^{2}    
u_{{\bol k}\lambda}(t)=0  ~~.    
\lab{p10a}   
\ee   
Eq. (\ref{p9}) (and (\ref{p10a})) is the familiar \cite{MTW} wave   
propagation equation for the vacuum gravitational field.  
When, instead of $\eta_{\mu\nu}$, the   
metrics   
\be   
g_{\mu \nu }^0=\left(     
\begin{tabular}{llll}    
1 & 0 & 0 & 0 \\     
0 & -$a^2$(t) & 0 & 0 \\     
0 & 0 & -$a^2$(t) & 0 \\     
0 & 0 & 0 & -$a^2$(t)    
\end{tabular}    
\right)     
\lab{p11}\ee   
with   
\be   
a(t)=a_0e^{\frac 13H t} ,   
\lab{p15}\ee   
and $H \,=\, const. = \frac{3\stackrel{.}{a}(t)}{a(t)} $   
(the Hubble constant),   
is adopted, $\overline{h}^{\mu \nu }$ may be still    
decomposed into partial waves   
as in (\ref{p10});    
however, in such a case one obtains the equation    
\cite{1, Grish1, Grish2, Grish3} (see also \cite{GG})   
\be   
\stackrel{..}{u}(t)+H \stackrel{.}{u}(t)+\omega ^2(t)u(t)=0     
\lab{p12}\ee   
with     
\be   
{{\omega}^2}(t)=\frac{k^{2}c^2}{a^2(t)}~~.   
\lab{p14}\ee   
In Eq.(\ref{p12}) we have used $u(t)\,\equiv\,u_{{\bol k}\lambda}(t)$.   
In the Minkowski   
space-time $\om$ is constant in time, but when the Universe expands,    
$\om$ is time-dependent, $\om = \om(t)$.    
   
The term $H \stackrel{.}{u}$ in eq. (\ref{p12}) is    
generally incorporated into the frequency term by using the    
conformal time variable $\eta$   
\cite{Grish1, Grish2, Grish3}; such a   
computational strategy is very    
useful in the phenomenological approach, however our purpose    
in this paper is to illustrate the subtleties of the canonical    
quantization for non-unitary    
time evolution and therefore we must explicitly take care of such    
a term in eq. (\ref{p12}). Only in this way the full structure of    
the state space will be revealed.   
   
It is interesting to remark that eq.(\ref{p12}) with $a(t)$,  $H$ and    
$\om(t)$ given by (\ref{p15}) and (\ref{p14})   
is also known as Hill-type equation \cite{Hill}. By setting $\xi = \ep \,    
z$, $\ep = \frac{3k c}{a_{0} H}$, $z= e^{-\frac{H}{3}t}$ and $u= z^{2} V$   
we can easily show that   
it can be written in the form of the spherical Bessel    
equation    
\be   
\xi^{2} V_{\xi\xi} + 2 \xi V_{\xi} + \left(\xi^{2}-2\right)V = 0   
~~. \lab{p17}\ee   
   
One particular solution of eq.(\ref{p17}) is indeed the spherical Bessel   
function of the first kind   
\be   
j_{1}(\xi)= \frac{\sin \xi}{\xi^{2}} - \frac{\cos \xi}{\xi}    
~~, \lab{p18}\ee   
i.e.   
\be   
u(t)=\frac{1}{\ep^2} \left[\sin(\ep e^{-\frac{H}{3}t})    
-  \ep e^{-\frac{H}{3}t}\, \cos(\ep e^{-\frac{H}{3}t})\right]    
~~,\lab{p19}\ee   
which in fact is solution of (\ref{p12}).

\setcounter{chapter}{3}   
\setcounter{equation}{0}   
\section*{3. Quantization in expanding geometry}   
   
We remark that the term  $H \stackrel{.}{u}$ in eq. (\ref{p12}) prevents  
any possibility to proceed to canonical quantization. As a matter of  
fact, it is not even possible to define the canonical conjugate momentum  
to the $u$ variable, $p_u=\frac{\partial    
L}{\partial \stackrel{.}{u}}$, since the system described by eq.  
(\ref{p12}) is a non-lagrangian one. 
  
However, according to   
\cite{QD, FT, Bat}, if one {\it doubles} the  
degrees of freedom of the system, then it becomes possible to 
recover the canonical formalism, as indeed we will see. Therefore  
we consider the double oscillator system   
\be\stackrel{..}{u}+H \stackrel{.}{u}+\omega ^2(t)u=  0 ~,\lab{p20}\ee    
\be \stackrel{..}{v}-H \stackrel{.}{v}+\omega ^2(t)v=  0 ~,   
\lab{p21}   
\ee   
and in this sense we may speak of a   
"double Universe". We observe indeed that   
in the same way eq. (\ref{p20}) is implied by the expanding (inflating)    
metrics, eq.   
(\ref{p21}) for the oscillator $v$ can be associated to the    
"contracting" ("deflating") metrics,   
complementary to the inflating one.    
   
In order to proceed in our discussion it is convenient to write down    
the Lagrangian in terms of $u$ and $v$ modes from which eqs.(\ref{p20}),    
(\ref{p21}) are directly obtained:   
\be L=\stackrel{.}{u}\stackrel{.}{v}+\frac 12H (u\stackrel{.}{v}-   
v\stackrel{.}{u})-\omega ^2(t)uv  ~.   
\lab{p30}\ee   
The canonical momenta can be {\it now} introduced    
\be p_u=\frac{\partial    
L}{\partial \stackrel{.}{u}}=\stackrel{.}{v}-\frac 12   
H v ~, \qquad \qquad   
 p_v=\frac{\partial L}{\partial    
\stackrel{.}{v}}=\stackrel{.}{u}+\frac 12H u ~,   
\lab{p32}\ee   
and the Hamiltonian is   
\be {\cal H} = p_{u} \dot u + p_{v} \dot v - L =  p_{u} p_{v} +    
{{1}\over{2}}   
H\left ( v p_{v} - u p_{u} \right ) + \left ( {\omega}^{2}(t) -{{H    
^{2}}\over{4}} \right ) u v \quad .  \lab{p33}\ee   
We put  
\be\Omega(t) \equiv \left [ \left ({\omega}^{2}(t)  - {{H    
^{2}}\over{4}} \right ) \right ]^{1\over{2}}, 
\lab{l1}\ee 
which we     
assume to be real, i.e. ${\omega}^{2}(t) > {{H ^{2}}\over{4}}$~ for any    
$t$: i.e. $k \geq   
{a_{0}H\over{2c}}e^{{H\over{3}}t}$. This tells us that as time flows   
the reality condition excludes long wave modes, which in turn acts as an   
intrinsic infrared cut-off.  
Further discussions on this point will be developed in Sect. 7. 
See also \cite{1}    
for a discussion of time domains in inflating models.    
   
It is interesting to observe that $v = u e^{Ht}$ is solution of (\ref{p21})   
and that by setting $u(t) = {1\over {\sqrt 2}} r(t)e^{{-Ht\over 2}}$ and    
$v(t) = {1\over {\sqrt 2}} r(t)e^{{Ht\over2}}$ the system of equations   
(\ref{p20}) and (\ref{p21}) is equivalent to the equation    
$   
\stackrel{..}{r} +\Omega ^2(t)r=  0 $,   
(see also \cite{BGPV}). This further    
clarifies the meaning of the doubling of the $u$    
oscillator: the $u-v$ system is a non-inflating (and non-deflating) system.    
This is why it is possible to set up the canonical quantization scheme   
for the doubled $u-v$ system.   
   
We introduce as usual the commutators   
\be[\, u ,    
p_{u} \, ] = i\, \hbar = [\, v , p_{v} \,] \quad , \quad  [\, u , v \,] = 0    
= [\, p_{u} , p_{v} \, ] ~~, \lab{p34}\ee   
and it turns out to be convenient to introduce the variables $U$    
and $V$ by    
the transformations   
\be U(t)\, =\, \frac{u(t) + v(t)}{\sqrt 2}, \qquad    
V(t)\, =\, \frac{u(t) - v(t)}{\sqrt 2}    
~~, \lab{p35}\ee   
which preserve the commutation relations (\ref{p34}):   
\be[\, U , p_{U} \, ] = i\, \hbar = [\, V , p_{V} \,] \quad ,    
\quad  [\, U ,    
V \,] = 0 = [\, p_{U} , p_{V} \, ] ~.   
\lab{p36}\ee   
   
In terms of the variables $U$ and $V$ it now appears that we are dealing    
with the decomposition of the parametric oscillator $r(t)$ on the {\it    
hyperbolic} plane (i.e. in the pseudo-Euclidean metrics): $r^{2}(t) =    
U^{2}(t) -V^{2}(t)$.   
    
We then proceed to the quantization (for more details see Appendix A)   
by introducing the annihilation and creation    
operators:   
\be   
A= \frac{1}{\sqrt{2}} \left(\frac{p_U}{\sqrt{\hbar\om_{0}}}-iU   
\sqrt{\frac{\om_{0}}{\hbar}}   
\right), \qquad\qquad   
A^{\dag}= \frac{1}{\sqrt{2}} \left(\frac{p_U}{\sqrt{\hbar\om_{0}}} + iU   
\sqrt{\frac{\om_{0}}{\hbar}}   
\right), \lab{p43}\ee   
\be   
B= \frac{1}{\sqrt{2}} \left(\frac{p_V}{\sqrt{\hbar\om_{0}}}-   
 iV\sqrt{\frac{\om_{0}}{\hbar}}   
\right), \qquad\qquad   
B^{\dag} = \frac{1}{\sqrt{2}} \left(\frac{p_V}{\sqrt{\hbar\om_{0}}} +    
 iV\sqrt{\frac{\om_{0}}{\hbar}}   
\right),   
\lab{p44}\ee   
with commutation relations   
\be   
[A,A^{\dag}] = 1 = [B,B^{\dag}], \quad [A,B] = 0, \quad    
[A^{\dag},B^{\dag}] = 0 ~,   
\lab{p45}\ee   
and all other commutators equal to zero. In eqs. (\ref{p43}), (\ref{p44})   
${\omega}_{0}$ denotes an arbitrary frequency \cite{Perel}.   
In terms of $A $ and   
$B $ the Hamiltonian is finally written as   
\be   
{\cal H}  = {\cal H}_{0} + {\cal H}_{I_1} + {\cal H}_{I_2}   
\lab{p58}\ee   
\be   
{\cal H}_{0} = \half \hbar \Om_{0 }(t)   
(A^{\dag}  A  - B^{\dag}  B  )   
\equiv \hbar \Om_{0 }(t){\cal C} ~,   
\lab{p59}\ee   
\be   
{\cal  H}_{I_1} = - {1\over 4}\hbar \Om_{1 }(t)   
\left[\left(A ^{2}   
+ {A^{\dag} }^{2} \right) - \left( B ^{2} + {B^{\dag}}^{2}    
\right)\right] ~,   
\lab{p60}\ee   
\be   
{\cal H}_{I_2} = i  {\Gamma}   
{\hbar} \left(A^{\dag}  B^{\dag}  -A B     
\right) \equiv i \hbar {\Gamma }   
( J_{+ } - J_{- }) ~.   
\lab{p61}\ee   
with ${\Gamma}  \equiv {{H}  \over 2}$. We note that for any t   
\be   
[{\cal H}_{0}, {\cal H}_{I_2}] = 0 = [{\cal H}_{I_1} , {\cal H}_{I_2}]~,   
\lab{p62}\ee   
which guarantees that under time evolution the minus sign appearing in    
${\cal H}_{0}$ is not   
harmful (i.e., once one starts with a positive definite Hamiltonian it    
remains   
lower bounded). We also note that the group structure underlying the   
Hamiltonian (\ref{p58}) is the one of SU(1,1) (see Appendix A).   
   
Finally, we recall that the $A$ and $B$ operators (and all other operators)    
as well as other quantities, e.g. ~${\omega}(t)$, actually    
are dependent on    
the momentum $\vec k$ (and on other degrees of freedom) and thus our    
formulas should be understood as carrying such a $\vec k$ labels which we    
have been omitting for simplicity.   
For instance the commutators (\ref{p45}) are indeed to be understood    
as   
\be   
[A_{\bol k},A^{\dag}_{{\bol k}'}] = {\delta}_{{\bol k},{\bol k}'} =    
[B_{\bol k},B^{\dag}_{{\bol k}'}], \quad [A_{\bol k},B_{{\bol k}'}] = 0,    
\quad [A^{\dag}_{\bol k},B^{\dag}_{{\bol k}'}] = 0 ~.   
\lab{p51}\ee   
   
Our next task is to study the Hilbert space structure and this will be done    
in the following section.

\setcounter{chapter}{4}   
\setcounter{equation}{0}   
\section*{4. The vacuum structure}

Again, for simplicity, we will omit the $\vec k$ indices whenever no    
misunderstanding arises.   
   
In order to study the eigenstates of the Hamiltonian (\ref{p58}), we    
consider the set  $\{ |n_{A} , n_{B} > \}$  of simultaneous   
eigenvectors of $A^{\dagger} A$ and $B^{\dagger} B$, with $n_{A}$, $n_{B}$   
non-negative integers.   
These are eigenstates of ${\cal H}_{0}$ with eigenvalues   
${1 \over 2}\hbar \Om_{0}(t)   
(n_{A} -n_{B})$ for any $t$.  The eigenstates of ${\cal H}_{I_2}$ can be    
written in the standard   
basis, in terms of the eigenstates of ${\left ( J_{3} -{1\over{2}} \right    
)}$ in the representation labeled by the value $j \in   
Z_{1\over{2}}$ of ${\cal C}$,  $\{ | j , m > \, ; \, m \geq |j| \}$: \be   
{\cal C} | j , m > = j | j , m > \quad , \quad j = {1\over{2}} (n_{A} -    
n_{B}) \quad ;   
\ee   
\be   
 \left ( J_{3} - {1\over{2}} \right ) | j , m > = m | j , m > \quad , \quad    
m = {1\over{2}} (n_{A} + n_{B}) \quad .  \lab{p71}   
\ee   
   
By using the $su(1,1)$ algebra   
\be   
[\, J_{+} , J_{-}\, ] = - 2 J_{3} \quad ,   
\quad [\, J_{3}  , J_{\pm}\, ] = \pm    
J_{\pm} \quad , \lab{p72}   
\ee   
one can show that   
the kets $| \psi_{j , m} > \equiv   
e^{\left ( + {{\pi}\over{2}} J_{1} \right )} | j , m >$   
satisfy indeed the equation $J_{2} | \psi_{j , m} > = \mu | \psi_{j , m}>$    
with pure imaginary $\mu \equiv i \left ( m + {1\over{2}} \right ) $   
(see Appendix B.1).   
   
We are left with the discussion of the eigenstates of  ${\cal H}_{I_{1}}$.    
We can "rotate away" \cite{so} ${\cal H}_{I_1}$   
 by using the transformation    
\be   
{\cal H} \rightarrow   
{{\cal H}^{\prime}} \equiv e^{i\theta(t) K_{2}}{\cal H}   
e^{-i\theta(t) K_{2}} = {\cal H}^{\prime}_{0} + {\cal H}_{I_{2}}~,    
\lab{p74a}\ee   
\be   
\tanh {\theta(t)} = -{\Om_1 (t) \over \Om_0 (t)}~,   
\lab{p74}\ee   
with   
\be   
{{\cal H}^{\prime}}_{0} = \hbar \Om(t)   
(A^{\dag} A - B^{\dag} B )   
\lab{p75}\ee   
and $[{{\cal H}^{\prime}}_{0} , {\cal H}_{I_{2}} ] = 0$.   
See the  
Appendix A for the definition of $K_{2}$ (eq.(\ref{p481})),   
${\Omega}_{0}$ and ${\Omega}_{1}$ (eq.(\ref{p46})).   
We have used the algebra (\ref{p53b}) and the fact    
that ${\cal H}_{I_2}$ commutes with   
$K_{2}$. Note that when the $\vec k$ indices are restored we   
have ${\theta(t)} \equiv {\theta}_{\bol k}(t)$. Also note that the choice   
$\tanh {\theta(t)} = -{\Om_1 (t) \over \Om_0 (t)}$ is allowed since the    
modulus of   
$(-{\Om_1 (t) \over \Om_0 (t)})$ is at most equal to one (for any $\vec k$    
and any $t$).   
   
In conclusion, the eigenstates of the Hamiltonian $\cal H$ at $t$, eq.   
(\ref{p58}), are states of type   
$e^{-i\theta(t) K_{2}} | \psi_{j , m} >$~:   
$$   
{\cal H}e^{-i\theta(t) K_{2}} | \psi_{j , m} > =   
e^{-i\theta(t) K_{2}} e^{i\theta(t) K_{2}}   
{\cal H}e^{-i\theta(t) K_{2} }| \psi_{j , m} > =   
$$   
\be   
e^{-i\theta(t) K_{2}}({\cal H}^{\prime}_{0} + {\cal H}_{I_2})   
|\psi_{j , m} > = \left( \hbar \Om (t) (n_A - n_B) -  i \hbar {\Gamma}(n_A +    
n_B + 1) \right)   
e^{-i\theta(t) K_{2} }| \psi_{j , m} > ~~,   
\lab{p76}\ee   
where we have been using the algebras (\ref{p53b}), (\ref{p72})   
and the commuting properties of the related Casimir operators. Note    
that $\Omega (t)$ appears to be the common frequency of    
the two oscillators    
$A$ and $B$. In conclusion, the spectrum of ${\cal H}$ is determined by    
using the so-called method   
of the spectrum generating algebras \cite{Perel, so}.   
   
The solution to the Schr\"odinger equation can be given with    
reference to the initial time pure state $| j , m_{0} >$ (see refs.    
\cite{QD,FT}). When in particular, the initial state, say at arbitrary    
initial time $t_{0}$, is the {\it vacuum} for ${{\cal H}^{\prime}}_{0}$,    
{\it i.e.} $| n_{A} = 0 , n_{B} = 0 > \equiv |0>$, with $A |0> = 0 =    
B|0>$ ({\it i.e.} $j=0 , m_{0}=0$ for any $\vec k$),   
the state    
\be   
|0(\theta(t_{0}))> = e^{-i\theta (t_{0}) K_{2} }|0> ~,   
\ee   
at $t_{0}$    
(and for given ${\vec k}$), is the zero energy eigenstate   
(the {\it vacuum}) of ${\cal H}_{0} + {\cal H}_{I_1}$ at $t_{0}$:   
\be   
({\cal H}_{0} + {\cal H}_{I_1})|_{t_{0}}|0(\theta (t_{0}))> =   
e^{-i\theta (t_{0}) K_{2} }{\cal H}^{\prime}_{0}|0> = 0 ~~    
for~~any~~arbitrary~~t_{0}.   
\lab{p77}\ee   
   
The state $|0(\theta (t))>$ is a generalized $su(1,1)$ squeezed   
coherent state, appearing, as well known, in the study of the parametric   
excitations of the quantum oscillator \cite{Perel}.   
   
In the following for simplicity we will put $t_{0} = 0$ and set $    
\theta(t_{0}=0) \equiv \theta$ and $|0(\theta (t_{0}=0))> \equiv   
|0(\theta)> $.   
   
Time evolution of the squeezed vacuum $|0(\theta)>$ is given by   
\be   
|0({\theta},t) > = \exp{\left ( - i t {{{\cal H}}\over{\hbar}}\right )}    
|0(\theta)> = \exp{\left ( - i t {{{\cal H}_{I_2}}\over{\hbar}}\right )}   
|0(\theta)> ~,   
\lab{p78}\ee   
due to (\ref{p62}) and to (\ref{p77}). Notice that in eq.(\ref{p78})    
${\cal H}$ denotes the Hamiltonian at time $t_{0}=0$, {\it i.e.} the    
Hamiltonian (\ref{p58})  with   
$\Omega_{0}(t_{0}=0)$ and $\Omega_{1}(t_{0}=0)$..     
   
We observe that the   
operators $A$ and $B$ transform under   
$\exp{\left ( -i\theta K_{2} \right )}$ (for any given ${\vec k}$) as   
\be   
A  \mapsto A (\theta) = {\it e}^{   
-i\theta K_{2} }A  {\it e}^{i\theta K_{2}}   
 =  A  \cosh{(\half\theta )} + A ^{\dagger} \sinh{(   
\half\theta )}~,   
\lab{p7261}\ee   
\be   
B  \mapsto B (\theta) = {\it e}^{   
-i\theta K_{2} }B  {\it e}^{i\theta K_{2} }   
 =  B  \cosh{(\half\theta )} + B ^{\dagger} \sinh{(   
\half\theta )}~.   
\lab{p7262}\ee   
These transformations   
are the well known squeezing    
transformations which preserve the commutation relations (\ref{p45}) (and    
(\ref{p51})).   
One has   
\be   
A (\theta) |0(\theta)> = 0 = B (\theta) |0(\theta)>    
\quad .   
 \lab{p27}\ee   
   
We have (at finite volume $V$)   
\be   
|0(\theta,t)> = {1\over{\cosh{(\Gamma  t)}}} \exp{   
\left ( \tanh {(\Gamma  t)} J_{+}(\theta) \right )} |0(\theta)> \quad ,    
\lab{p79}\ee   
with $J_{+}(\theta) \equiv A ^{\dagger}(\theta) B^{\dagger}(\theta)$,    
namely a generalized $su(1,1)$ squeezed coherent state   
with equal numbers of modes $A (\theta)$ and   
$B (\theta)$    
condensed in it (for each $\vec k$) at each $t$.   
We observe that   
\be   
<0(\theta,t) | 0(\theta,t)> = 1 \quad \forall t \quad , \lab{p710}\ee   
\be   
<0(\theta,t) | 0(\theta)> = \exp{\left ( - \ln \cosh {(\Gamma  t)} \right    
)} \quad ; \lab{p711}\ee   
which shows how, provided ${\Gamma > 0}$ ,   
\be   
<0(\theta,t) | 0(\theta)> \, \propto \exp{\left ( - t  \Gamma \right )}   
\rightarrow 0 \; ~for~large~t .   
\lab{p712}\ee   
   
Thus eq. (\ref{p712}) shows the vacuum instability:    
time evolution brings out of the initial-time Hilbert space for large $t$.    
This is not acceptable in quantum mechanics since there the Von Neumann    
theorem states that all the representations of the canonical commutation    
relations are unitarily equivalent and therefore there is no room in    
quantum mechanics for   
non-unitary time evolution as the one in (\ref{p712}).   
On the contrary, in QFT there exist infinitely many unitarily inequivalent    
representations and this suggests to us to consider    
our problem in the framework of QFT, which we will do in the next section.

\setcounter{chapter}{5}   
\setcounter{equation}{0}   
\section*{5. Quantum field theory in expanding geometry}

To set up    
the formalism in QFT we have to consider the infinite volume limit;    
however, as customary, we will work at finite volume and at the end of    
the computations we take the limit $V \rightarrow \infty$. The QFT    
Hamiltonian is introduced as   
\be   
{\cal H}  = {\cal H}_{0} + {\cal H}_{I_1} + {\cal H}_{I_2},   
\lab{ph1}\ee   
\be   
{\cal H}_{0} =  \sum_{\bol k} \half \hbar \Om_{0,{\bol k}}(t)   
(A^{\dag}_{\bol k} A_{\bol k} - B^{\dag}_{\bol k} B_{\bol k} )   
= \sum_{\bol k} \hbar \Om_{0,{\bol k}}(t){\cal C}_{\bol k} \equiv   
\sum_{\bol k} \hbar \Om_{0,{\bol k}}(t) K_{0,{\bol k}} ~,   
\lab{ph2}\ee   
\be   
{\cal  H}_{I_1} = - \sum _{\bol k}{1\over 4}\hbar \Om_{1,{\bol k}}(t)   
\left[\left(A_{\bol k}^{2}   
+ {A^{\dag}_{\bol k}}^{2} \right) - \left( B_{\bol k}^{2} + {B^{\dag}_{\bol    
k}}^{2} \right)\right]   
\equiv - \sum_{\bol k} \hbar \Om_{1,{\bol k}}(t)K_{1,{\bol k}} ~,   
\lab{ph3}\ee   
\be   
{\cal H}_{I_2} = i\sum _{\bol k} {\Gamma}_{\bol k}   
{\hbar} \left(A^{\dag}_{\bol k} B^{\dag}_{\bol k} -A_{\bol k}B_{\bol k}    
\right) \equiv i\sum_{\bol k} \hbar {\Gamma_{\bol k}}   
( J_{+,{\bol k}} - J_{-,{\bol k}}) ~.   
\lab{ph4}\ee   
   
 We also have now   
\be   
{{\cal H}^{\prime}}_{0} = \sum_{\bol k}\hbar \Om_{\bol k}(t)   
(A_{\bol k}^{\dag} A_{\bol k} - B_{\bol k}^{\dag} B_{\bol k} )~~,   
\lab{ph5}\ee   
and, at finite volume $V$,    
\be   
|0(\theta,t)> = \prod_{\bol k} {1\over{\cosh{(\Gamma_{\bol k} t)}}} \exp{   
\left ( \tanh {(\Gamma_{\bol k} t)} J_{{\bol k}, +}(\theta) \right )}    
|0(\theta)> \quad , \lab{pq79}\ee   
which corresponds to eq.(\ref{p79}).    
The state $|0({\theta},t)>$ is also a    
$su(1,1)$ squeezed coherent state. Eqs. (\ref{p710})-(\ref{p712}) are    
now replaced by   
\be   
<0(\theta,t) | 0(\theta,t)> = 1 \quad \forall t \quad , \lab{pq710}\ee   
\be   
<0(\theta,t) | 0(\theta)> = \exp{\left ( - \sum_{\bol k} \ln \cosh    
{(\Gamma_{\bol k} t)} \right )} \quad , \lab{pq711}\ee   
which again exhibits non-unitary time evolution, provided ${\sum_{\bol k}    
\Gamma_{\bol k} > 0}$:   
\be   
<0(\theta,t) | 0(\theta)> \, \propto \exp{\left ( - t  \sum_{\bol k}     
\Gamma_{\bol k} \right )} \rightarrow  0 \; ~ for~large~t~ . \lab{pq712}\ee   
Use of the customary   
continuous limit relation $ \sum_{\bol k}    
\mapsto {V\over{(2 \pi)^{3}}} \int \! d^{3}{{\vec k}}$,    
for $\int \!   
d^{3}{\vec k} \, \ln \cosh {(\Gamma_{\bol k} t)}$    
finite and positive, gives in the   
infinite volume limit   
\be   
<0(\theta,t) | 0(\theta)> \mapbelow{V \rightarrow \infty} 0 \qquad\qquad \forall    
\, t \quad ,   
\lab{p713}\ee   
\be   
<0(\theta,t) | 0(\theta',t') > \mapbelow{V \rightarrow \infty} 0 \quad    
with ~~ \theta' \equiv \theta (t_{0}'),~~\forall \, t\, , t'\, , t_{0}'    
\quad , \quad t \neq t'~~ .  \lab{p714}\ee   
   
We conclude that in the infinite volume limit, vacua at $t$ and at $t'$,   
~$\forall \, t ~ ,~ t' $ ~, with ~~ $t \neq t'$~, are orthogonal and the    
corresponding Hilbert spaces are unitarily inequivalent representations   
of the canonical commutation relations \cite{QD} (see also \cite{DFV}).   
   
Under time evolution generated by ${\cal H}_{I_{2}}$ the    
operators $A_{\bol k}(\theta)$ and   
$B_{\bol k}(\theta)$ transform as   
\be   
A_{\bol k}(\theta) \mapsto A_{\bol k}(\theta,t) = {\it e}^{- i    
{t\over{\hbar}} {\cal H}_{I_2}} A_{\bol k}(\theta) {\it e}^{i    
{t\over{\hbar}} {\cal H}_{I_2}} =  A_{\bol k}(\theta) \cosh{(\Gamma_{\bol k}    
t)} - B_{\bol k}^{\dagger}(\theta) \sinh{(   
\Gamma_{\bol k} t)}~,   
\lab{p7261a}\ee   
\be B_{\bol k}(\theta) \mapsto B_{\bol k}(\theta,t) = {\it e}^{- i    
{t\over{\hbar}} {\cal H}_{I_2}} B_{\bol k}(\theta)   
{\it e}^{i {t\over{\hbar}}    
{\cal H}_{I_2}} =  - A_{\bol k}^{\dagger}(\theta) \sinh{(\Gamma_{\bol k} t)} +    
B_{\bol k}(\theta) \cosh{(   
\Gamma_{\bol k} t)} ~.   
\lab{p7262a}\ee   
See the Appendix B.2 for the    
relation between the ${\Gamma}_{\bol k}$'s and  ${\Gamma}    
\equiv {{H} \over 2}$. Eqs. (\ref{p7261a}) and (\ref{p7262a})   
are Bogoliubov transformations and they can be    
understood as inner auto-morphism for the algebra $su_{\bol k} (1,1)$   
preserving   
the commutation relations (\ref{p51}).   
Thus, at every $t$ we have a copy   
$\{ A_{\bol k}(\theta,t),$ $A_{\bol k}^{\dagger}(\theta,t) ,   
B_{\bol k}(\theta,t) ,   
B_{\bol k}^{\dagger}(\theta,t) \, ; \, | 0(\theta,t) >\,   
|\, \forall {\vec k} \}$ of the   
original algebra and of its highest weight vector   
$\{ A_{\bol k}(\theta), A_{\bol k}^{\dagger}(\theta) , B_{\bol k}(\theta)    
, B_{\bol k}^{\dagger}   
(\theta) \, ; \, | 0(\theta) >\, |\,   
\forall {\vec k} \}$, induced by the time evolution operator, i.e. we have a   
realization of the operator algebra at each time t ( which can be    
implemented by Gel'fand-Naimark-Segal construction in the $C^{*}$-algebra    
formalism \cite{Bratteli, Ojima}).    
The time evolution operator therefore acts as a    
generator   
of the group of auto-morphisms of $\bigoplus_{\bol k}   
su_{\bol k} (1,1)$ parameterized by time $t$.  We stress that   
the copies of the original algebra provide unitarily inequivalent    
representations of the canonical commutation relations in the   
infinite-volume limit, as   
shown by eqs. (\ref{p714}).   
   
Also note  
that the operators ${K_{i,{\bol k}}}$,   
$i = 0,1,2$ (or $i= 0,+,-$), for fixed ${\vec k}$ close the algebra    
$su_{\bol k}(1,1)$ and that they commute for any $i,j$ for ${\vec k} \neq   
{\vec k}'$ (Appendix A).  
   
At each time $t$ one has   
\be   
A_{\bol k}(\theta,t) |0(\theta,t)> = 0 = B_{\bol k}(\theta,t)   
|0(\theta,t)> \quad,   
\quad \forall    
\, t \quad ,   
 \lab{p273}\ee   
and the number of modes of type $A_{\bol k}(\theta)$ in the state    
$|0(\theta,t)>$ is given, at each instant $t$ by   
\be   
{n}_{A_{\bol k}}(t) \equiv < 0(\theta,t) | A_{\bol k}^{\dagger}   
(\theta)    
A_{\bol k}(\theta) | 0(\theta,t) > =   
 \sinh^{2}\bigl( \Gamma_{\bol k} t \bigr) \quad ,   
\lab{p274}\ee   
and similarly for the modes of type $B_{\bol k}(\theta)$.    
   
We also observe that the commutativity of ${\cal C}$ (i.e. $K_{0}$)    
with ${\cal    
H}_{I_2}$ (i.e. $J_{2}$) ensures that the number $\left( n_{A_{\bol k}} - n_{B_{\bol k}}    
\right)\,$ is a constant of motion for any $\vec k$ and any $\theta$.   
Moreover, one can show \cite{QD,TFD} that   
the creation of a mode $A_{\bol k}(\theta)$ is equivalent to the    
destruction of a mode $B_{\bol k}(\theta)$ and vice-versa.  This means    
that the $B_{\bol k}(\theta)$ modes can be interpreted  as the {\sl    
holes} for the modes $A_{\bol k}(\theta)$: the $B$-system can be    
considered as the heat bath for    
the $A$-system.   
   
The results obtained    
in this section (see also Appendix B) clearly show the r\"ole of    
the term  $H \stackrel{.}{u}$    
and its interplay with the time-dependent frequency term in   
eq. (\ref{p12}). By using the conformal time coordinate as usually done   
in the literature we would not be able to reveal the underlying rich   
structure of the state space.    
Such a structure naturally leads us to recognize the    
thermal properties of   
$|0({\theta},t)>$. This will done in the following section.   
   
\setcounter{chapter}{6}   
\setcounter{equation}{0}   
\section*{6. Entropy and free energy in expanding Universe}   
   
The vacuum state $|0({\theta},t)>$ as given by equation    
(\ref{pq79}) can be    
written as   
\be   
|0({\theta},t)>\, = \exp{\left ( - {1\over{2}} {\cal S}_{A({\theta})}    
\right )} |\,{\cal I}({\theta})>\, = \exp{\left ( - {1\over{2}} {\cal    
S}_{B({\theta})} \right )} |\,{\cal I}({\theta})> \quad ,   
\lab{p81}\ee   
where   
\be   
|\,{\cal I}({\theta})>\, \equiv \exp {\left( \sum_{\bol k}   
A_{\bol k}^{\dagger}({\theta})   
B_{\bol k}^{\dagger}({\theta}) \right)} |0({\theta})> ~,   
\lab{p82}\ee   
and   
\be   
{\cal S}_{A({\theta})} \equiv - \sum_{\bol k} \Bigl \{   
A_{\bol k}^{\dagger}({\theta}) A_{\bol k}({\theta})    
\ln \sinh^{2} \bigl ( \Gamma_{\bol k} t \bigr ) - A_{\bol k}({\theta})    
A_{\bol k}^{\dagger}({\theta}) \ln \cosh^{2} \bigl ( \Gamma_{\bol k} t   
\bigr ) \Bigr \} ~.   
\lab{p83}\ee   
   
${\cal S}_{B({\theta})}$ is given by the same expression    
with $B_{\bol k}   
({\theta})$ and $B_{\bol k}^{\dagger}({\theta})$   
replacing $A_{\bol k}(\theta)$ and   
$A_{\bol k}^{\dagger}({\theta})$,   
respectively.  In the following we    
shall simply write ${\cal S}({\theta})$ for either ${\cal    
S}_{A({\theta})}$   
or ${\cal S}_{B({\theta})}$.   
${\cal S}({\theta})$ is recognized to be the entropy (see    
Appendix B). Also note that   
$<0(\theta,t)| {\cal S}({\theta}) |0(\theta,t)>$   
grows monotonically with $t$: the entropy (for both $A$ and $B$)   
increases as the system evolves in time.  
   
Moreover, the difference ~~${\cal    
S}_{A(\theta)} - {\cal S}_{B(\theta)}$   
is constant in time (cfr. (\ref{p281})):   
\be   
[\, {\cal S}_{A(\theta)} - {\cal S}_{B(\theta)} , {\cal H}    
] = 0 \quad .   
\lab{p88}\ee   
   
${\cal S}_{A(\theta)} -   
{\cal S}_{B(\theta)}$~~ is in fact the (conserved) entropy for the closed   
system.   
   
Eqs. (\ref{p81}) and (\ref{p83}) show that    
time evolution is expressed solely in terms of the    
(sub)system $A$ ($B$)   
with the elimination of the (bath) $B$ ($A$) variables.    
   
At finite volume $V$, we obtain   
\be   
{{\partial}\over{\partial t}} |0(\theta ,t)> =  -  {1\over{2}} \left (    
{{\partial {\cal S}({\theta})}\over{\partial t}} \right )   
|0(\theta ,t)>  \quad .   
\lab{p91}\ee   
   
Thus $ {1\over{2}} \left (    
{{\partial {\cal S}({\theta})}\over{\partial t}}  \right ) $ is the   
generator of time-translations: time    
evolution is controlled by the entropy variations.   
  
This is a remarkable result: the operator that   
controls time evolution is the variation of the dynamical variable   
whose expectation value is formally an entropy.   
These features reflect indeed correctly the   
irreversibility of time evolution characteristic of expanding geometry.   
Expanding metrics implies in fact breaking of   
time-reversal invariance, i.e. the choice of a privileged   
direction in time evolution ({\it time arrow}).  
 
We have thus   
obtained, as a distinct feature of the canonical formalism, that   
the arrow of time in which entropy increases (the thermodynamic   
arrow) is in accordance with the arrow of time in which geometry   
expands (the cosmological arrow). This is in agreement with   
Penrose and Hawking \cite{Penrose, Hawking, book, Elitzur}  
semi-classical discussion on arrows of time.  
We will come back on this point and we will see how the Hawking  
boundary conditions are also satisfied in our formalism.  
   
In conclusion our result is that    
the system in its evolution runs over a variety of    
representations of the canonical commutation relations which are unitarily    
inequivalent to each other for $t \neq t'$ in the infinite-volume limit:    
{\it the non-unitary character of time evolution implied by expanding  
geometry is thus     
recovered,  in a consistent QFT scheme, in the unitary in-equivalence among    
representations at different times in the infinite-volume limit}.   
   
The above analysis    
suggests to us that the doubling of the degrees of freedom is intimately    
related with the non-trivial metric structure of space-time, the doubled    
degree of freedom signaling the lost of the Poincar\'e invariance.   
   
We want now to further analyze the thermal concepts and properties of the    
formalism above presented. For the sake of   
definiteness, let us consider the $A$-modes alone and introduce the    
functional   
\be   
F_{A} \equiv <0({\theta},t)| \Bigl ( {{\cal H}^{\prime} }_{0,A(\theta) } -   
 {1\over{\beta}} {\cal S}_{A({\theta})} \Bigr ) |0({\theta},t)> ~.   
\lab{p92}\ee   
Here ${{\cal H}^{\prime}}_{0,A(\theta)} \equiv \sum_{\bol k} E_{\bol k}    
A_{\bol k}^{\dagger}(\theta) A_{\bol k}(\theta)$;   
$\beta$ is a strictly    
positive function of time to be determined and $E_{\bol k} \equiv   
\hbar \Omega_{\bol k}(t_{0}=0) - \mu$, with $\mu$ the chemical potential.   
   
We write ${\sigma}_{\bol k} \equiv \Gamma_{\bol k} t$, and look for the    
values of $\sigma_{\bol k}$ making ${F}_{A(\theta)}$ stationary:   
\be   
{{\partial {F}_{A(\theta)}}\over{\partial \sigma_{\bol k}}} = 0 \quad ; \quad    
\forall \vec k \quad .   
\lab{p93}\ee   
   
Condition (\ref{p93}) is a stability condition to be satisfied for    
each representation. We now assume that $\beta$ is a slowly varying    
functions of t and thus eq. (\ref{p93}) gives   
\be   
\beta(t) E_{\bol k} = - \ln \tanh^{2} \bigl ( \sigma_{\bol k} \bigr ) \quad ,   
\lab{p94}\ee   
which together with (\ref{p274}) leads to  
\be   
{n}_{A_{\bol k}}(t) = \sinh^{2} \bigl ( \Gamma_{\bol k} t \bigr ) =    
{1\over{{\rm e}^{\beta (t) E_{\bol k}} - 1}} \quad , \lab{p101}   
\ee   
i.e.the Bose distribution for $A_{\bol k}$ at time $t$ provided   
we assume $\beta (t)$ to represent the inverse temperature    
$\beta(t) = {1\over{k_{B} T(t)}}$ at time $t$ ($k_{B}$    
denotes the Boltzmann   
constant).     
$\{ |0(\theta ,t)> \}$ is thus recognized to be a    
representation of   
the canonical commutation relations at finite temperature, equivalent   
with the Thermo Field Dynamics representation $\{ |0(\beta )> \}$ of    
Takahashi and Umezawa ~\cite{TFD} .    
   
In conclusion we can interpret ${F}_{A}$ as the free energy and    
${n}_{A}$ as the average number of $A$-modes at the inverse temperature    
$\beta (t)$ at time $t$.   
   
The change in time of the energy ${E_{A} \equiv \sum_{\bol k} E_{\bol k}    
{n}_{A_{\bol k}}}$ is given by   
\be   
d E_{A} ={{\partial}\over{\partial t}} \Bigl ( <0(\theta, t)| {\cal    
H}^{\prime}_{0,A(\theta)} |0(\theta, t)> \Bigr ) d t  =    
\sum_{\bol k} E_{\bol k}    
\dot{n}_{A_{\bol k}}(t) d t \quad ; \lab{p981}\ee   
and the change in the entropy by   
\be   
d S_{A} = {{\partial}\over{\partial t}} \Bigl (    
<0(\theta,t)| {\cal S}_{A(\theta)} |0(\theta,t)>    
\Bigr ) = \beta \sum_{\bol k} E_{\bol k}     
\dot{n}_{A_{\bol k}}(t) d t  = \beta d E_{A}(t) \quad ,   
\lab{p982}\ee   
provided we assume the changes in time of $\beta$ can be neglected   
(which happens, {\it e.g.} in the case of adiabatic variations of    
temperature, at $T$ high enough) . Thus   
we have   
\be   
d E_{A} - {1\over{\beta}} d {S}_{A} = 0 \quad ,   
\lab{p983}\ee   
consistently with the relation   
obtained directly by minimizing the free energy   
\be   
d {F}_{A} = d E_{A} - {1\over{\beta}} d {S}_{A} \quad =0 ~~.   
\lab{p984}\ee   
Eq. (\ref{p984}) expresses the   
first principle of thermodynamics for a system coupled with environment   
at constant temperature and in absence of mechanical work.   
$E_{A}$ is thus recognized to be the internal energy of the system.   
The above discussion shows that time evolution produces transitions    
over inequivalent representations by inducing changes in the number of    
condensed modes in the vacuum. By defining as usual heat as   
${dQ={1\over{\beta}} dS}$ we see that the change in time    
$\dot{n}_{A}$ of particles condensed in the vacuum turns out into heat    
dissipation $dQ$ \cite{QD}. Finally, (\ref{p984}) also shows    
that, provided    
variations of $E_{A}$ in the temperature are negligible, entropy is as    
usual the free energy response to temperature variations.   
   
The thermodynamics above discussed   
implies that the at early times in the expansion, when the system is  
"small", it is in a state of higher order (lower entropy) with   
respect to later times in the expansion, when it is "larger" and   
changes in the condensate turn out into heat dissipation   
and disordering (higher entropy). We thus recover in a consistent   
scheme the Hawking boundary conditions \cite{Hawking}  
to be satisfied in order that the   
thermodynamic arrow agrees with the cosmological one.

\setcounter{chapter}{7}   
\setcounter{equation}{0}   
\section*{7. Finite life-time for $k$-modes}   
  
In this section we analyze the time evolution of modes of   
specific momentum $k$. We will see that we can define a life-time   
for each $k$-mode and that, as already observed in Sect. 3,  
an intrinsic infrared cut-off is contained in the formalism. 
 
The reality condition for    
$\Omega_{k}(t)$ (cf. Sect. 3),  
such a condition actually excludes long wave modes, and thus acts as an  
intrinsic infrared cut-off. In fact, it is easy to show that   
\be {\Omega_{k}}^{2}(t) \ge 0 \Longrightarrow  k 
\ge k_{0} e^{{{H \over {3}}t}}, 
\lab{l2}\ee  
with $ k_{0} \equiv {Ha_{0} \over  {2c}}$ 
for  any positive $t$. Of course, this is welcome for the  
well-definiteness of the operator fields. Moreover,   
we recover in this way  
the known feature of inflationary models by which only in   
the ``tight coupling'' phase  
($\lambda \,<\, R_{H}$) there is an oscillatory evolution\cite{Albr}.  
     
We observe now that, besides the reality condition,   
we also have the condition (\ref{p74})  which implies that   
${\Omega_{k}}^{2}(t) \le {\omega_{0}}^{2}$ for $\theta_{k} \ge 0$ and that  
${\Omega_{k}}^{2}(t) \ge {\omega_{0}}^{2}$ for $\theta_{k} \le 0$.   
This, together with the reality condition, leads to the bounds for   
$k$:   

  \be  
  k_{0} e^{{H \over {3}}t} \le k \le {\tilde k}_{0}    
  e^{{H \over {3}}t}  
    ~~~ at~~~ any~~~t~~~for~~~  
  {\theta}_{k} \ge 0~,  
  \lab{c2}\ee   
  \be  
   k \ge {\tilde k}_{0}    
  e^{{H \over {3}}t}  
    ~~~ at~~~ any~~~t ~~~for~~~  
  {\theta}_{k} \le 0~,  
  \lab{c4}\ee   
  with ${\tilde k}_{0} \equiv   
   \frac{a_{0}}{c}\,\sqrt{\omega^{2}_{0} +   
  \left(\frac{H}{2}\right)^{2}}~$.  
 
  On the other hand, {\it for each given mode $k$}, the frequency   
  $\Omega_{k}$ is different from zero only in the span of time   
  $0 \le t \le {\rm T}_{k} \equiv   
  {3 \over {H}}\ln {k \over {k_{0}}}$   
  (limiting ourselves to positive time evolution).   
   
This result may be made more  
transparent by introducing the variable $\Lambda_{k}(t)\,\equiv\, 
  \theta_{k}(t)\ -\ \theta_{k}(0)$, i.e. (cf. eq.(\ref{p74})) 
  \be  
   e^{-2\Lambda_{k}(t)} \ =\   
\frac{e^{-t \frac{H}{3}}\,{\rm sinh}\frac{H}{3}({\rm T}_{k}-t)}{{\rm sinh}  
  \frac{H}{3}{\rm T}_{k}}, \qquad {\rm with}\,~  
\Lambda_{k}(t) \ \ge 0, \,{\rm for \, any \, t} ~,  
\lab{c6}\ee  
and rewriting, for ${\theta}_{k} \ge 0$, eq.(\ref{l1}) in the form 
  \be {\Omega_{k}(\Lambda_{k}(t))}\,=\,  \Omega_{k}(0) e^{-\Lambda_{k}(t)}~,  
  \lab{c5}\ee  
  with ${\Omega_{k}(\Lambda_{k}(0))}\,=\,  \Omega_{k}(0) $  and  
  ${\Omega_{k}(\Lambda_{k}({\rm T}_{k}))}\,=\,0$.   
  %
   

    
   
  \begin{figure}[t] 
  \caption{``Lives'' of $k$-modes, for growing $k$} 
  \epsfig{file=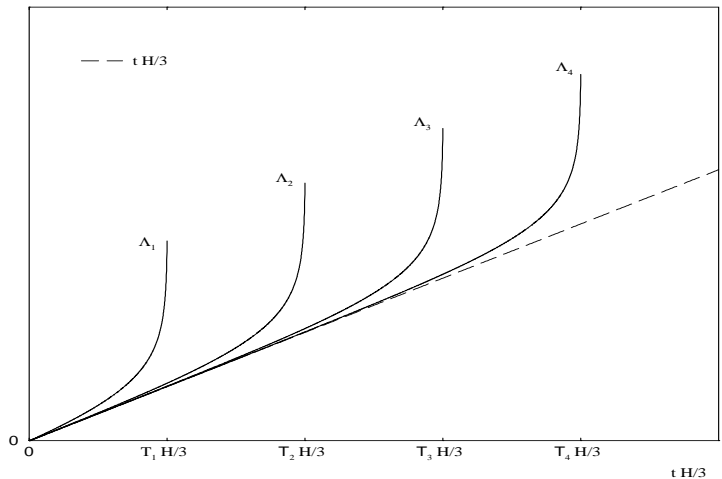,width=0.8\linewidth,height=0.5\linewidth} 
  \end{figure} 

Furthermore, eq. (\ref{c6}) shows that $\Lambda_{k}(0)=0$,  
  $\Lambda_{k}(T_{k})\rightarrow\infty$, suggesting to interpret  
  $\Lambda_{k}$ as the  life-time of the $k$-mode, i.e.  
$\Lambda_{k}(t) =  \propto  \tau_{k}(t)$ 
denoting     
$\tau_{k}(t)$ the {\it proper} time of the $k$-mode. 
 
Modes with larger $k$ have "longer" life with reference  
to time $t$ (see figure).  
   
In other words, each $k$-mode "lives" with a proper time $\tau_{k}$, so  
when $\tau_{k}$ is zero, the mode is born, dying for $\tau_{k}$  
$\rightarrow$ $\infty$. In figure several "lives" are shown, for  
growing $k$; $\Lambda_{k}$'s are drawn versus $tH/3$, reaching the  
blowing up values in correspondence of the abscissa points $T_{k}H/3$. 
 
In conclusion, only the modes 
satisfying conditions (\ref{c2}) and (\ref{c4}) are present at time $t$, 
being the other ones decayed.  
 
It is finally interesting to remark that the number $n_{k}$ of   
$k$-modes condensed in the state $|0(\theta)>$, given by   
$\sinh^{2}(\theta_{k})$, can be expressed as   
${n_{k}\,\equiv\,n_{+,k}+n_{-,k}}$ 
in terms of the $k$-modes   
$n_{+,k}$ and $n_{-,k}$   
going forward and backwards in the proper time $\tau_{k}$,  
respectively: 
  \be  
  n_{\pm,k}(t)=\frac{n_k(t)}{  
  \ e^{\mp 2 \theta_{k}(0)}   
  e^{\mp 2 \Lambda_{k}(t)} +1} 
  \ee  
 
Clearly, one recognizes that   
$n_{+,k} - n_{-,k}$ can be considered as an "order parameter"  
since   
$\tanh{\theta_{k}(t)} = -  \frac{n_{+,k}(t) -   
  n_{-,k}(t)}{n_{k}(t)}$~, and $n_{+,k} = n_{-,k}$ for any $t$  
for each $k$-mode with constant frequency ($\theta_{k} = 0$   
for any $t$).  
 
Let us close this section with two comments. 
 
The infrared cut-off implied by the reality condition  
for $\Omega_{k}(t)$ means that infinite volume contributions are  
missing, and this in turn means that transitions through states  
at different $t$'s $|0(\theta,t)>$ are possible  
(cf. eq. (\ref{pq712}); at infinite volume eq. (\ref{p714}) holds). 
 
Finally, we observe that the shorter life-time of modes of lower  
$k$ (long wave-lenghts) and the longer survival of modes of higher $k$  
(shorter wave-lenghts) might signal the formation in later times of  
inhomogeneities (again in agreement with Hawking boundary  
conditions \cite{Hawking}) whose size is of the order of the wave  
lenghts of the survived modes. Moreover, this might also be an  
interesting mechanism able to explain energetic relic bursts  
possibly generated by high momentum gravitational modes. 
Work on this subject is in progress.

\setcounter{chapter}{8}   
\setcounter{equation}{0}   
\section*{8. Final remarks and conclusions}   
   
In this paper we have studied    
the vacuum structure for expanding Universe.   
We have considered only the gravitational    
wave modes in the FRW metrics in a de Sitter phase.     
   
We have shown that the system state space splits into many unitarily    
inequivalent representations of the canonical commutation relations    
parameterized by time $t$ and non-unitary time    
evolution is    
then described as a trajectory in the space of the representations: the    
system evolves in time by running over unitarily inequivalent    
representations. This means that the full set of unitarily    
inequivalent Hilbert spaces must be exploited, which provides further    
support to known results in QFT in curved space-time \cite{MSV, WA1,    
WA2}. The generator of time evolution is related to the entropy operator,    
which indeed reflects the irreversibility in time evolution ({\it the    
arrow of time}).   
At the same time, entropy appears as the response of    
free energy to the temperature variation and thus the intrinsic thermal    
character of the non-unitary time evolution is exhibited, also    
recovering results of ref. \cite{Prok}   
(see also \cite{Kim2}). The    
vacuum turns out to be the generalized SU(1,1) squeezed   
coherent state of thermo    
field dynamics, thus exhibiting the link between    
non-unitary time evolution    
in expanding geometry, squeezing and thermal properties.  
   
Hawking and Penrose \cite{Penrose, Hawking, book},   
although from different perspectives, have shown that solutions   
which start out with an inflationary period, will have a well   
defined thermodynamic arrow. In particular, Hawking,  
on the basis of semi-classical arguments, has discussed   
the boundary conditions to be satisfied in order for the   
thermodynamic arrow and the cosmological arrow of time to agree   
in both the expanding and the contracting phases. These   
conditions imply that the Universe is in a smooth state of higher   
order when it is small but that it may be in an inhomogeneous    
disordered state when it is large \cite{Hawking}.  
We remark that the conclusions   
we have reached in the canonical formalism fully agree with Penrose and   
Hawking conclusions and they are also consistent with Hawking   
boundary conditions. In particular we note that the thermodynamic   
nature of the vacuum turns out in our discussion to be a specific   
feature of the inflationary time evolution, thus confirming   
Hawking predictions. Furthermore, Hawking conclusion 
that "if there is a   
constant probability of the Universe expanding, there must be an   
equal probability of it contracting" \cite{Hawking}   
finds its counter-part in our formalism in the fact that the   
contracting geometry is the "mirror in time" image of the   
expanding geometry (cf. eqs. (\ref{p20})-(\ref{p21})).  
 
Finally, in the case of   
black hole solutions, provided one considers the doubled variable   
as the one inside the black hole (as for example in \cite{MSV}), our   
discussion also agree with Hawking prediction that the   
thermodynamic arrow should be reversed inside black holes (cf.   
e.g. eq. (\ref{p88})). For what concerns Hawking psychological arrow   
of time,  
it is interesting to observe that there exist a quantum model of   
memory \cite{QDB} where memory processes are shown to point also   
in the same time direction of the thermodynamic arrow. But we   
will not insist more on this point. 
  
It is an interesting question whether    
the results on the scalar field in an expanding geometry    
of refs. \cite{Guth, Eboli, Allen},   
the self-similar time-dependent scale transformation for the functional    
Schr\"odinger equation for a free scalar field \cite{Vaut} and the    
effective potential models for inflating Universe \cite{Boya}   
can also be incorporated in the present scheme.   
   
The doubling of the    
degrees of freedom has revealed to be    
a fundamental tool in our discussion.  
It appears to be the bridge to the    
unified picture of non-unitary time evolution, squeezing and thermal    
properties in expanding metrics.   
In this paper we have clarified its physical meaning    
and have  shown that it is not simply a mathematical device:   
{}From the thermal properties perspective the    
physical interpretation of the    
doubled degrees of freedom is the one of the thermal bath degrees of    
freedom; from the point of view of the vacuum structure the one of {\it    
holes} of the relic gravitons; from the Hamiltonian formalism point of    
view the one of the {\it complement} to the inflating system.   
   
As a    
matter of fact, the doubling of the degrees of freedom is    
intrinsic to the    
Bogoliubov transformations, so that one deals with a doubled    
system anytime    
one works with such transformations.   
For this reason all the "mixed modes" formalisms (since    
Parker's work \cite{Park} ) necessarily involve the algebraic structure    
of the doubling of the modes.   
   
For example, we remark that in ref. \cite{Prok}  
the doubling of the degrees of freedom is actually introduced by    
considering the modes of momentum $\bf k$ and $- \bf k$ as the {\it    
couple} of modes of total zero momentum in terms of which the two-mode    
squeezed vacuum is described: the distinction between the   
$\bf k$ and $- \bf k$ modes introduces a partition in the $\bf k$ space    
and leaves out the zero momentum modes which,    
although not entering in the condensate structures (in \cite{Prok} the    
summations are always limited to ${\bf k} > 0$), nevertheless are present in    
the quantized field $\phi$ and in its canonical momentum $\pi$.   
Also in \cite{Grish1, Grish2, Grish3} the doubling is actually present.   
In fact,    
in considering the general solution of the parametric oscillator $u(t)$   
the authors introduce    
{\it two} different representations of $u(t)$ in terms of two distinct    
basis, $\xi(\eta)$ and $\chi(\eta)$, respectively, with $\eta$   
playing the role of time coordinate. In this way the doubling of the    
degrees of freedom is introduced.

As a further remark we would like to point out that    
the "negative" kinetic    
term in our Lagrangian structure (cf. section 3) also appears    
in two-dimensional gravity models in conjunction    
with the problem of the ambiguity of the vacuum definition \cite{Jack}.  
In the case where no negative norm states appear in the theory,  
its canonical structure    
is similar to the one of    
non-unitary time evolution described in the present paper.

\section*{Acknowledgments}   
We are grateful to Professor L.P.Grishchuk for an interesting discussion.   
We are also grateful to the anonymous referee who has suggested   
that it could be useful for us    
to comment on the semi-classical discussion of the arrows of   
time by Penrose and Hawking. 
  
This work has been partially supported by INFN, by MURST and by the  
European Science Foundation Network on Topological defects   
formation in phase transitions.  
   
\newpage   
   
\appendix   
\setcounter{chapter}{1}   
\setcounter{equation}{0}

\section*{ Appendix A}   
   
The Lagrangian (\ref{p30}) may be rewritten in terms of $U$ and $V$ as   
\be   
L= L_{0,U} - L_{0,V}  + {H \over 2}({\dot U} V - {\dot V} U) ~~, \lab{p37}\ee   
with   
\be L_{0,U} =  {1\over 2} \dot U^2 - {{\omega^2}(t)\over 2} U^2 ~~,   
~~~L_{0,V} =  {1\over 2} \dot V^2 - {{\omega^2}(t)\over 2} V^2 ~~.   
\lab{p38}\ee   
The associate momenta are   
\be   
p_U =  \dot U + {H \over 2} V ~~,\qquad   
p_V = - \dot V -{H \over 2} U ~~, \lab{p39}\ee   
and the motion equations corresponding to the set (\ref{p20})-   
(\ref{p21}) are   
\be   
 \ddot U + H \dot V + {\omega^2}(t) U  = 0  ~~,\lab{p40-a}\ee   
\be   
 \ddot V + H \dot U + {\omega^2}(t) V  = 0  ~~. \lab{p40-b}\ee   
   
The Hamiltonian (\ref{p33}) becomes   
\be   
{\cal H} = {\cal H}_U - {\cal H}_V = {1 \over 2} (p_U - {H \over 2}V)^2   
+ {{\omega^2 (t)}\over 2} U^2 - {1 \over 2} (p_V + {H \over 2}U)^2 -   
{{\omega^2 (t)}\over 2} V^2 ~~. \lab{p41}\ee   
   
Eq.(\ref{p37}) shows that the $H$-term actually acts as a    
coupling between the oscillators $U$ and $V$ and produces a    
correction to    
the kinetic energy for both oscillators (cf.eq.(\ref{p41})).   
The group structure underlying the   
Hamiltonian (\ref{p41}) is the one of SU(1,1).   
By putting $x = U, V$, we indeed have    
\be   
{1 \over 2} {p_x}^2 + {1 \over 2}{\Omega}^2(t)x^2 =   
{\Om}_{0} K_{0}^{x} - {\Om}_{1} K_{1}^{x}   
\lab{p45c}\ee   
with   
\be   
\Omega_{0,1}= \om_{0} \left(\frac{\Om^{2}(t)}{\om^{2}_{0}}\pm 1\right) ~~,   
\lab{p46}\ee   
and   
\be   
{K_{0,1}}^x = \frac{1}{2 \om_{0}}\left(\frac{p_{x}^{2}}{2} \pm    
\frac{x^{2}}{2}   
\om^{2}_{0}   
\right) ~,   
\lab{p48}\ee   
which together with   
\be   
{K_{2}}^x = \frac{1}{4}(p_{x}x+xp_{x})   
\lab{p481}\ee   
close the $su(1,1)$ algebra:   
\be   
[K_{1}^x,K_{2}^x]=-i K_{0}^x, \quad [K_{2}^x,K_{0}^x]=i K_{1}^x, \quad   
[K_{0}^x,K_{1}^x]= i K_{2}^x ~.   
\lab{p53}\ee   
   
Of course,   
$[K_{i}^x ,K_{j}^{x'}] = 0$ for any $i,j$ and $x \neq x'$.   
By using eqs. (\ref{p43})-(\ref{p44}) and eqs. (\ref{p48}) and    
(\ref{p481}) we introduce then the operators   
$$   
K_{0} = \half (A^{\dag}A -B^{\dag} B )    
\qquad \qquad   
K_{1 }= \qrt \left[\left(A ^{2}   
+ {A^{\dag} }^{2} \right)  - \left(B ^{2}   
 + {B^{\dag} }^{2}  \right)\right] , $$   
\be K_{2 }= i\qrt \left[ \left(A ^{2}   
 - {A^{\dag} }^{2}  \right) + \left( B ^{2}   
 -{B^{\dag} }^{2}  \right)\right] , \lab{p55}\ee   
which also close the algebra $su(1,1)$:   
\be   
[K_{1},K_{2}]=-i K_{0}, \quad [K_{2},K_{0}]=i K_{1}, \quad   
[K_{0},K_{1}]= i K_{2}~.   
\lab{p53b}\ee   
We further remark that   
\be   
J_{+ }= A^{\dag} B^{\dag} , \quad  J_{- }=    
A B , \quad   
J_{0 }= \frac{1}{2}(A^{\dag} A +   
B^{\dag} B  + 1),   
\lab{p56}\ee   
with $J_{1 } = \half (J_{+ } + J_{- })$ and   
$J_{2 } = -{i \over 2} (J_{+ } - J_{- })$ ,   
also close the $su(1,1)$ algebra. Note that   
\be   
2iJ_{2 } =   
  A^{\dag} B^{\dag}  - A B   
\lab{p561}\ee   
commutes with each $K_{i} ,~i=0,1,2$ (eqs. (\ref{p55})), and that   
\be   
{\cal C}  \equiv \frac{1}{2}(A^{\dag} A  -B^{\dag}B ) = {K_{0 }}   
~~ \lab{p57}\ee   
commutes with each $J_{i} ,~i=0,+,-$ (eqs. (\ref{p56})). $2iJ_{2}$ and    
$\cal C$ are indeed (related to) the Casimir operators for the algebras of    
generators (\ref{p55}) and (\ref{p56}), respectively.   
   
\appendix   
\setcounter{chapter}{2}   
\setcounter{section}{0}   
\setcounter{equation}{0}

\section*{ Appendix B }   
   
\section{}   
In order to find the eigenvalues for the Hamiltonian (\ref{p58}),    
we have to   
analyze the $J_{2}$ spectrum.   
   
Notice that although $J_{2}$ appears to be hermitian, it has   
pure imaginary discrete spectrum in $| \psi_{j , m} >$. This apparent    
contradiction is related with the   
well known \cite{FT, agi} fact that the states $| \psi_{j , m} >$ are not    
normalizable and   
the transformation generating $| \psi_{j , m} >$ from $|j , m >$ is a   
non-unitary transformation in $SU (1,1)$ (it is not a proper rotation in    
$SU(1,1)$, but is rather   
a pseudo-rotation in $SL ( 2 , C )$ \cite{QD}).  In other words   
 $|\psi_{j , m} >$ does not   
provide a unitary irreducible representation ({\sl UIR}), consistently     
with the fact \cite{ln} that in any {\sl UIR}   
of $SU(1,1)$ $J_{2}$ should have purely continuous and real spectrum    
(which does not happen in the case of $| \psi_{j , m} >$). However,   
{\sl UIR} with continuous and real spectrum are not adequate for the    
description of inflating phenomena and therefore they are of no help to    
us.   
   
We can bypass this difficulty by    
introducing in the Hilbert   
space a new metric with a suitable inner product in such a    
way that   
$| \psi_{j ,m} >$ has a finite norm \cite{QD,FT}  .    
We consider therefore the   
anti-unitary operation ${\cal T}$ under which   
$(A , B) \mapright{\cal T} ( - A^{\dagger} , -B^{\dagger})$ and    
introduce the conjugation   
$< \psi_{j , m} | \equiv$ $[\, {\cal T} | \psi_{j    
 , m} >\, ]^{\dagger}$, with   
${\cal T} | \psi_{j , m} > \equiv$ $| \psi_{j , - (m+1)} >$.   
The hermitian of   
$J_{2} | \psi_{j , m} > = \mu | \psi_{j , m}> ~,~   
\mu \equiv i \left ( m +    
{1\over{2}} \right )$, is now    
\be   
< \psi_{j , - (m+1)} | J_{2} = \mu_{\cal T} < \psi_{j , - (m+1)} | \quad ,    
\quad \mu_{\cal T} = - i \left [\, - (m+1) + {1\over{2}}\, \right ] =    
 - \mu^{*} = \mu\quad .   
\lab{p73}\ee   
   
With reference to the states  $|0>$ and $|0(\theta (t_{0})>$ we   
notice that ${\cal H}^{\prime}_{0}|0> = 0$ for any $t$, but that   
$({\cal H}_{0} + {\cal H}_{I_1})|0> \neq 0$ and ${\cal H}^{\prime}_{0}    
|0(\theta (t_{0}))> \neq 0$ for any $t$ (cf. (\ref{p77})) . However,   
expectation values of ${\cal H}^{\prime}_{0}$ and of   
$({\cal H}_{0} + {\cal H}_{I_1})$ in $|0>$ and    
in $|0(\theta (t_{0}))>$ are   
all zero at any $t$.   
   
We observe that    
the number of modes of type $A $ in the state $|0(\theta)>$   
is given,    
by   
\be   
n_{A }(\theta) \,\equiv\,    
< 0(\theta) | A ^{\dagger} A  |    
0(\theta) > \,=\, \sinh^{2}\bigl( \theta  \bigr) \quad ;   
\lab{p272}\ee   
and similarly for the modes of type $B $.   
Moreover, the commutativity of $J_2$ with $K_{2}$ and    
(\ref{p7261})-(\ref{p7262}) ensure that   
\be   
 A^{\dag} B^{\dag}  - A B  =   
  A^{\dag} ({\theta})B^{\dag} ({\theta}) - A({\theta})B ({\theta})~.    
 \lab{p28}\ee   
Thus   
$$   
{\cal H}_{0} + {\cal H}_{I_{1}} + {\cal H}_{I_{2}} =   
e^{-i\theta K_{2}}{{\cal H}^{\prime}}_{0}e^{i\theta K_{2}} +   
{\cal H}_{I_{2}} =$$   
\be   
 = \hbar \Om(0)   
(A^{\dag}(\theta) A(\theta) - B^{\dag}(\theta) B(\theta) ) +    
i\hbar \Gamma   
( A^{\dag} ({\theta})B^{\dag} ({\theta}) - A({\theta})B ({\theta}))~,    
\lab{p281}\ee   
   
Finally, in the QFT framework   
$|0(\theta)>$ is explicitly given by   
\be   
|0(\theta)> = \prod_{\bol k} {1\over{\cosh{(\theta_{\bol k})}}} \exp{   
\left ( \tanh {(\theta_{\bol k})} j_{{\bol k}, +} \right )} |0> \quad ,    
\lab{p792}\ee   
with $j_{{\bol k} ,+} \equiv a_{\bol k}^{\dagger} b_{\bol    
k}^{\dagger} = {1\over2}({A_{\bol k}^{\dagger}}^{2} + {B_{\bol    
k}^{\dagger}}^{2})$, and $a_{\bol k} = {1\over{\sqrt 2}}(A_{\bol k} +   
iB_{\bol k})$, $b_{\bol k} = {1\over{\sqrt 2}}(A_{\bol k} - iB_{\bol    
k})$. Thus, $|0(\theta)>$ also is a $su(1,1)$  generalized    
coherent state.    
Its time evolution is given by   
\be   
|0({\theta},t)> = \prod_{{\bol k},{\bol q}} {1\over{\cosh{(\Gamma_{\bol k}    
t)}}{\cosh{(\theta_{\bol q})}} } \exp{   
\left ( \tanh {(\Gamma_{\bol k} t)} J_{{\bol k}, +}(\theta) \right )}    
\exp{   
\left ( \tanh {(\theta_{\bol q})} j_{{\bol q}, +} \right )} |0> \quad ,    
\lab{p793}\ee   
or, alternatively, by   
\be   
|0({\theta},t)> = \prod_{{\bol k},{\bol q}} {1\over{\cosh{(\Gamma_{\bol k}    
t)}}{\cosh{(\theta_{\bol q})}} } \exp{   
\left ( \tanh {(\theta_{\bol q})} j_{{\bol q}, +}(t) \right )}   
\exp{   
\left ( \tanh {(\Gamma_{\bol k} t)} J_{{\bol k}, +} \right )} |0> \quad .    
 \lab{p794}\ee   
where $J_{{\bol k}, +} = A_{\bol k}^{\dagger} B_{\bol    
k}^{\dagger}$, ~ $j_{{\bol q}, +}(t) \equiv   
a_{\bol k}^{\dagger}(t) b_{\bol k}^{\dagger}(t) = {1\over2}({A_{\bol    
k}^{\dagger}}^{2}(t) + {B_{\bol k}^{\dagger}}^{2}(t))$, ~  $a_{\bol    
k}(t) = {1\over{\sqrt 2}}(A_{\bol k}(t) + iB_{\bol k}(t))$, $b_{\bol    
k}(t) = {1\over{\sqrt 2}}(A_{\bol k}(t) - iB_{\bol k}(t))$, ~    
$[a_{\bol k}(t), a_{\bol q}^{\dagger}(t)] = {\delta}_{{\bol    
k},{\bol q}} = [b_{\bol k}(t), b_{\bol q}^{\dagger}(t)]$ and all other    
commutators equal to zero. The operators $A_{\bol k}(t)$ and $B_{\bol    
k}(t)$ are given by the (canonical) Bogoliubov transformations   
\be   
A_{\bol k} \mapsto A_{\bol k}(t) = {\it e}^{- i {t\over{\hbar}} {\cal    
H}_{I_2}} A_{\bol k} {\it e}^{i {t\over{\hbar}} {\cal H}_{I_2}} =  A_{\bol    
k} \cosh{(\Gamma_{\bol k} t)} - B_{\bol k}^{\dagger} \sinh{(   
\Gamma_{\bol k} t)} \quad ,   
\lab{795}\ee   
\be   
B_{\bol k} \mapsto B_{\bol k}(t) = {\it e}^{- i {t\over{\hbar}} {\cal    
H}_{I_2}} B_{\bol k} {\it e}^{i {t\over{\hbar}} {\cal H}_{I_2}} =  -    
A_{\bol k}^{\dagger} \sinh{(\Gamma_{\bol k} t)} + B_{\bol k} \cosh{(   
\Gamma_{\bol k} t)} \quad .   
\lab{p796}\ee

\section{}   
   
In the continuum limit eqs. (\ref{p51})   
become   
\be   
[A_{\bol k},A^{\dag}_{{\bol k}'}] = {\delta}({\bol k} - {\bol k}') =    
[B_{\bol k},B^{\dag}_{{\bol k}'}], \quad [A_{\bol k},B_{{\bol k}'}] = 0,    
\quad [A^{\dag}_{\bol k},B^{\dag}_{{\bol k}'}] = 0 ~.   
\lab{p517}\ee   
and, as well known, the $A_{\bol k}$ (and $B_{\bol k}$) operators are    
not well defined on vectors in the Fock space; for instance $|A_{\bol k}>    
\equiv A_{\bol k}^{\dag}|0>$ is not a normalizable vector since from eqs.    
(\ref{p517}) one obtains $<A_{\bol k}|A_{\bol k}> = \delta ({\bol 0})$    
which is infinity.   
As customary one must then introduce wave-packet (smeared out) operators    
$A_{f}$ with spatial distribution described by square-integrable    
(orthonormal) functions   
\be   
f(\bol x) = {1 \over{(2\pi)^{3}}}\int {d^3{\bol k}} f({\bol k}) e^{i{\bol    
k}{\bol x}}   
\lab{p798}\ee   
i.e.   
\be   
A_{f} = {1 \over{(2\pi)^{3/2}}}\int {d^3{\bol k}} A_{\bol k} f({\bol k})   
\lab{p797}\ee   
with commutators   
\be   
[A_{f},A^{\dag}_{g}] = (f,g) = [B_{f},B^{\dag}_{g}],   
 \quad [A_{f},B_{g}] = 0, \quad [A^{\dag}_{f},B^{\dag}_{g}] = 0 ~,   
\lab{p518}\ee   
with $(f,g)$ denoting the scalar product between $f$ and $g$. Now    
$<A_{f}|A_{f}> = 1$ and the $A_{f}$'s are well defined operators   
in the Fock space in terms of which the observables have to be   
expressed. In this connection it is interesting to recall that the   
reality condition on $\Omega(t)$ (see sect. 3) naturally introduces the   
infrared cut-off smearing out the operator fields. In conclusion, we   
express the number operator as   
\be   
{n}_{A_{f}}(t) \equiv < 0(\theta,t) | A_{f}^{\dagger}(\theta) 
A_{f}(\theta)    
| 0(\theta,t) > =   
 {1 \over{(2\pi)^{3}}}\int {d^3{\bol k}}   
\sinh^{2} \bigl( \Gamma_{\bol k} t \bigr) |f({\bol k})|^{2}   
\equiv \sinh^{2} \bigl( \Gamma t \bigr) ~ ,   
\lab{p2746}\ee   
and similarly for the modes of type $B_{f}(\theta)$ (cf. with    
eq. (\ref{p274})). Eq. (\ref{p2746}) specifies the relation between the    
${\Gamma}_{\bol k}$'s and ${\Gamma}$ and says that the number of $A_{f}$    
modes does not depend on the volume.   
   
\section{}   
    
{}From eq. (\ref{p81}) we obtain    
\be   
|0({\theta},t)> = \sum_{n \geq 0} \sqrt{W_{n}(t)}\, | n(\theta) ,    
n(\theta) > \quad ,   
\lab{p84}\ee   
where $n(\theta)$ denotes the multi-index $\{ n_{\bol k}(\theta) \}$, and   
\be   
W_{n}(t) = \left ( \prod_{\bol k} {{\cosh^{2(n_{\bol k}+1)} \bigl (    
\Gamma_{\bol k} t \bigr )}\over{\sinh^{2 n_{\bol k}} \bigl ( \Gamma_{\bol k} t    
\bigr )}} \right )^{-1} \quad , \quad 0 < W_{n} < 1 \quad .     
\lab{p85}\ee   
Notice that the expansion (\ref{p84}) contains only terms for which    
$n_{A_{\bol k}({\theta})}$ equals $n_{B_{\bol k}({\theta})}$   
for all $\vec k$'s and that   
\be   
\sum_{n \geq 0} W_{n}(t) = 1 \quad \quad {\rm for\; any}\; t,   
\lab{p86}\ee   
\be   
<0(\theta,t)| {\cal S}({\theta}) |0(\theta,t)> =   
- \sum_{n \geq 0} W_{n}(t)    
\ln{W_{n}(t)} \quad . \lab{p87}\ee   
Eq. (\ref{p87}) leads us therefore to interpreting   
${\cal S}({\theta})$ as the {\it    
entropy} \cite{QD, TFD}.   
We also observe that  $<0(\theta,t)| {\cal S}({\theta}) |0(\theta,t)>$   
grows monotonically    
with $t$: the entropy for both $A$ and $B$   
increases as the system evolves in time.



\begin{thebibliography}{99}  
  
  
\bibitem{1} Brandenberger R H 1995 {\em Rev. Mod. Phys.} {\bf 57} 1   
  
\bibitem{GU} Guth A H 1981 {\em Phys. Rev. } {\bf D23} 347  
  
\bibitem{2} Calzetta E and Hu B L  1995 {\em Phys. Rev. } {\bf D52} 6770  
  
\bibitem{Park} Parker L 1969 {\em Phys. Rev.} {\bf D183} 1057   
  
\bibitem{5} Hawking S W  1975 {\em Comm. Math. Phys.} {\bf 43} 199   
  
\bibitem{Grish1} Grishchuk L P 1974 {\em Zh. Eksp. Teor. Fiz.} {\bf 67}  
825 [1975 {\em Sov. Phys. JEPT} {\bf 40} 409]  
\bibitem{Grish2} Grishchuk L P  and Sidorov Y V 1990 {\PRD}{42} 3413    
and refs. quoted therein  
\bibitem{Grish3} Grishchuk L P and Haus H A  and Bergman K    
\Journal {1992} {\PRD}{46} 1440   
\bibitem{MSV} Martellini M, Sodano P and Vitiello G 1978 
{\it Nuovo Cimento} {\bf 48 A} 341   
%
%
\bibitem{WA1} Wald R M {\it Preprint gr-qc/9509057} Talk given at the 14-th Int. Conf. on  
General Relativity and Gravitation (GR/4), Florence 6-12 August 1995  
  
\bibitem{WA2} Wald R M 1994 {\em Quantum Field Theory in Curved Space and   
Black Hole Thermodynamics} (Chicago: University of Chicago)  
 
\bibitem{Guth} Guth A and Pi S Y 1985 {\it Phys. Rev.} {\bf D32} 1899   
  
\bibitem{Eboli} Eboli O, Jackiw R and Pi S Y 1988 {\it Phys. Rev.} {\bf   
D37} 3557 \\ 
Eboli O, Pi S Y and Samiullah M 1989 {\it Ann. Phys.}{\bf 193} 102 
  
\bibitem{Allen} Allen B 1985 {\it Phys. Rev.} {\bf D32} 3136   
  
\bibitem{DFV} De Filippo S and Vitiello G 1977 {\it Lett.   
Nuovo Cimento} {\bf 19} 92   
  
\bibitem{QD} Celeghini E, Rasetti M and Vitiello G 1992    
{\em Ann. Phys.}{\bf 215} 156    
  
\bibitem{Penrose} Penrose R 1968 {\em Structure of the Spacetime} 
eds. DeWitt  and Wheeler\\ 
Hawking S W  1971 {\em Phys. Rev. Lett.} {\bf 26} 1344\\ 
Hawking S W 1972 {\em Comm. Math. Phys.} {\bf 27} 283 
  
\bibitem{Hawking} Hawking S W 1985 {\em Phys. Rev.} {\bf D32} 2389 
 
\bibitem{book} Hawking S W, Penrose R 1996  
{\em The Nature of Space and Time} 
(Princeton University Press) 
 
\bibitem{Elitzur} Elitzur A, Dolev S 1999 {\em Phys. Lett.}  
{\bf A251} 89 
 
\bibitem{TFD} Takahashi Y and Umezawa H 1975  {\it Collective
Phenomena}{\bf 2} 55   

\bibitem{Um1} Umezawa H, Matsumoto H  and Tachiki M  1982 {\em Thermo Field   
Dynamics and Condensed States} (Amsterdam: North-Holland Pub. Co.)  
 
\bibitem{Um} Umezawa H 1993 {\em Advanced Field Theory: Micro, Macro, and   
Thermal Concepts} (New York: American Institute of Physics)  
  
\bibitem{Israel} Israel W 1976 {\em Phys. Lett.} {\bf 57A} 107   
  
\bibitem{Jo} Johansson A E I, Umezawa H and Yamanaka Y 1990 {\em Class.   
Quantum Grav.} {\bf 7} 385   
  
\bibitem{SW} Sewell G L, 1982 {\em Ann. Phys. (N.Y.)} {\bf 141} 201  
  
\bibitem{Prok} Prokopec T 1993 {\em Class. Quant. Grav.}  
 {\bf 10} 2295  
  
\bibitem{BraProk} Brandenberger R H, Mukhanov V  and Prokopec T   
1993 {\em Phys. Rev.} {\bf D48} 2443   
  
\bibitem{Albr} Albrecht A, Ferreira P, Joyce M and Prokopec T,     
\Journal{1994}{\PRD}{ 50} {4807}\\  
 Albrecht A, Coherence and   
Sakharov Oscillations in the Microwave Sky,   
{\em Preprint astro-ph 9612015.}     

\bibitem{MTW}Missner W, Thorne K S, Wheeler J A, 1973 {\it Gravitation}, 
(New York:Freeman), chap. 35. 

\bibitem{GG} Gasperini M and Giovannini M  1993 {\it Phys. Rev.} {\bf
D47} 1519   
  
\bibitem{Hill} Profilo G and Soliani G 1991 {\it Progr. Theor. Phys.}{
\bf 84}  974 (1990); {\it Phys. Rev.} {\bf A44} 2057 (1991)\\  
 Profilo G and Soliani G 1994 {\it Ann. Phys.   
(N.Y.)} {\bf 229} 160   
  
\bibitem{FT} Feshbach H and Tikochinsky Y 1977 {\it
Transact. N.Y. Acad. Sci.}  
{\bf 38} (Ser. II) 44   
  
\bibitem{Bat}  Bateman H 1931 {\it Phys. Rev.} {\bf 38}  815  
  
  
\bibitem{BGPV} Blasone M, Graziano E, Pashaev O K and Vitiello G 1996  
{\it Ann. Phys. (N.Y.)} {\bf 252} 115   
  
\bibitem{Perel} Perelomov A M {\it Generalized Coherent States and   
thei Applications} (Springer-Verlag, Berlin 1986)  
  
\bibitem{so} Solomon A I 1971 {\it J. Math. Phys.} {\bf 12} 390   
  
\bibitem{Bratteli} Bratteli O and Robinson D W 1979 {\it Operator  
Algebras and Quantum   
Statistical Mechanics} (Berlin: Springer) 
  
\bibitem{Ojima} Ojima I 1981 {\it Ann. Phys. (N.Y.)} {\bf 137} 1   
 
\bibitem{Kim2} Kim S P and Kim S W 1995 {\it Phys. Rev.} {\bf D51}   
4254  
 
\bibitem{QDB} Vitiello G 1995, {\em Int. J. Mod. Phys.} {\bf B9} 973\\ 
              Alfinito E, Vitiello G, {\it in preparation} 
 
\bibitem{Vaut} Braghin F L, Martin C and  Vautherin D 1995   
{\it Phys. Lett.} {\bf B348} 343   
  
\bibitem{Boya} Boyanowski D, De Vega H J    
and Holman R 1994 {\it Phys. Rev.}{\bf D49} 2769   
  
\bibitem{Jack} Cangemi D, Jackiw R and   
Zwiebach B 1996  {\it Ann. Phys. (N.Y.)}{\bf 245} 408   
  
\bibitem{agi} Alhassid Y,  G\"ursey F and   
Iachello F 1983 {\it Ann. Phys.}  
({\it N.Y.}) {\bf 148} 346   
  
\bibitem{ln} Lindblad G and Nagel B 1970   
{\it Ann. Inst. H. Poincar\'e} {\bf XIII A} 27    

\end{thebibliography}
\end{document}